# Exploiting Conceptual Knowledge for Querying Information Systems

Joachim Selke and Wolf-Tilo Balke

Institute for Information Systems, Technische Universität Braunschweig, Germany

**Abstract.** Whereas today's information systems are well-equipped for efficient query handling, their strict mathematical foundations hamper their use for everyday tasks. In daily life, people expect information to be offered in a personalized and focused way. But currently, personalization in digital systems still only takes explicit knowledge into account and does not yet process conceptual information often naturally implied by users. We discuss how to bridge the gap between users and today's systems, building on results from cognitive psychology.

## 1. Introduction

Given today's information overload querying digital information in a personalized manner has become a more and more demanding problem. To a large degree this information is stored in structured repositories and databases allowing for efficient retrieval. Commonly used query languages require users to explicitly specify hard constraints to describe their information needs. In most cases however, this is extremely difficult for the user, since information needs are often vague and not all relevant parameters are known a priori, i.e., before interacting with the information system. In fact, much knowledge is embodied in the user or perceived as being common knowledge (Balke and Mainzer 2005). Moreover, when evaluating such queries over a repository or database, very often empty results are returned in case of overspecified queries, or the user is flooded by irrelevant information when posing underspecified queries. Recently, these problems have stimulated research in the area of personalized query processing. But still only few approaches actually focus on the cognitive processes underlying human information search.

A first step to human-centered information search are preference-based retrieval models, which build on simple pairwise item comparisons, in the form of 'items with property A are better than those showing property B.' This kind of statement is very intuitive for users and easy to articulate. Characteristic properties of items in a collection usually can be described within a fixed attribute scheme. For example, used cars on sale can be described in terms of brand, model, age, color, or top speed. Users now are generally able to specify their preferences on attribute level like 'I like VWs better than Ferraris' or 'I prefer red cars to yellow ones.' Often such characteristic attributes are simple and intuitive representations of concepts that would be hard to describe otherwise. Considering the sample statements above, the user might have an understanding that VWs are typical 'normal' cars, whereas Ferraris are typical examples of sports cars. On the other hand, the above color preference does not carry any information about other properties a user might want for his/her car (Leslie, 2007; Leslie, 2008). Obviously, integrating notions of typicality and user expectations into query processing would heavily improve the result quality in information searches as well as the user's search experience. Currently, no information system is able to process such implicit knowledge. In this paper we discuss the ingredients needed for designing information systems shaped for personalized and intuitive querying.

## 2. Overview of Current Preference-Based Query Processing

Current preference-based information systems rely on classical preference models, which expose strong mathematical properties. These properties are intended to capture rational behavior being expressed as internal coherence and logical consistency within a system of beliefs and preferences (e.g., requiring preference relations to be transitive and constant over time, contexts, and occasions); choices are consequential, and options are evaluated using prior assessments of beliefs and values (Mellers, Schwartz, & Cooke, 1998).

There are two major approaches in today's systems: utility functions (Fishburn, 1968; Fishburn, 1970; Keeney & Raiffa, 1976) and relational preference structures (Öztürk, Tsoukiàs, & Vincke, 2005; Kießling, 2002; Chomicki, 2003). A utility function assigns a numerical score to each data item. The score of an item depends only on the item's properties and represents its overall desirability. Relational preference structures link pairs of items through the notions of 'is preferred to' and 'is equally preferable as,' thus leading to qualitative preference orderings. Note that every utility function implies some simple-structured preference ordering. The problem of actually building personalized preference-based information systems has been investigated in various settings such as querying databases, getting automated product recommendations, or receiving support in decision making.

During the 1990s, the top-k retrieval paradigm has been introduced into database research to make systems more cooperative (Fagin, 1999; Güntzer, Balke, & Kießling, 2000; Hwang & Chang, 2007). Here, scores for item characteristics are aggregated by a suitable user-provided utility function. The resulting preference ordering can be computed quite efficiently and only the best items are returned to the user. Thus, the result of a query is neither empty, nor unmanageably large. But providing an adequate aggregation function is often difficult and not intuitive for the user, limiting the paradigms usefulness. In contrast, the skyline query paradigm does not rely on a user-provided function, but incorporates user preferences with respect to individual item characteristics using the notion of Pareto optimality as known from economics (Börzsönyi, Kossmann, & Stocker, 2001; Godfrey, Shipley, & Gryz, 2007). Again, result sets never can become empty but often are unmanageably large due to its over-cautious approach to the elimination of items.

Recommender systems (Adomavicius & Tuzhilin, 2005) target a different problem setting. Given a set of previous item ratings the system derives a personalized set of recommendations for each user. Currently, the most successful approaches rely on collaborative filtering, where suggested items are selected based on ratings of similar users (Candillier, Meyer, & Boullé, 2007). Although recommendations are easy to obtain, the paradigm falls short in case the users did not yet provide sufficiently enough ratings, the so-called cold-start problem. Furthermore, the preference order cannot be explicitly adjusted to the individual information need.

For the application in decision support techniques from the area of multi-criteria decision making are used. After a thorough decision analysis, the problem can be represented within a formal model. There is a large variety of techniques to fuse conflicting criteria, eventually deriving a set of 'optimal' solutions, e.g., by estimating criterion weights within linear utility models. Since quality of decisions is hard to measure in practice, the discipline focuses on designing methods to be as 'rational' as possible. This follows the idea that a rational decision process will always arrive at the best solution given the current knowledge. A related approach for eliciting and analyzing user preferences is conjoint analysis, which is a standard tool in market research (Green, Krieger, & Wind, 2001).

Generic approaches to preference handling can be found in the area of artificial intelligence (Boutilier, Brafman, Domshlak, Hoos, & Poole, 2004; Doyle, 2004; McGeachie & Doyle, 2004; Domshlak & Joachims, 2007). Reasoning frameworks infer additional knowledge from basic preference statements within a well-defined logic. Moreover, machine learning techniques can be employed to derive complete preference structures from small sets of preference statements. Most of these approaches however stay at a theoretical level, where the efficiency needed for actual system building is not a primary issue. Instead, methods are designed without a specific application scenario in mind and try to build complete models of domain knowledge. However, large parts of these models are not beneficial for answering actually occurring user queries. Furthermore, complicated explanations generated from long implication chains following a logical calculus are hardly useful in practical problem solving.

A different approach to automated reasoning is case-based reasoning (Lenz, Bartsch-Spörl, Burkhard, & Wess, 1998). Here, new problem instances (so-called cases) are solved by exploiting and transferring knowledge gained from similar, previously experienced cases. In contrast to most traditional reasoning frameworks, case-based reasoning focuses on solving only the current problem instance. It avoids broad generalization and does not generate an explicit model of the problem domain. Although this enables computationally efficient problem solving, explaining the motives for suggesting a specific solution to the user is difficult. In particular, it is hardly possible for the user to derive an intuition about the problem domain from the suggested solutions.

In summary, there is a wide variety of systems. However, in practical applications they are not intuitively useable. In all types of systems, the query formulation requires a large effort, often beyond the user's capabilities. As we will see, most intuitive preference statements carry implicit information that either is considered obvious and not worth mentioning, or is extremely hard to state in crisp terms, or the user is entirely unaware of. Whereas databases and decision support systems require comprehensive and explicit preference specification, recommender systems gain their predictive power only from implicit preference statements such as item ratings. In neither case the user is really at ease with the systems as practical experiments show (Pu & Chen, 2007). We will argue that cognitively supportive systems have to work in the middle of both extremes.

## 3. A Cognitive Psychology's Guide to Preferential Choice

Looking at results from cognitive psychology we get a quite different picture of human preferences than reflected in today's information systems. Empirical studies of choice processes revealed surprising but systematic deviations from predictions made by classical preference models. In the following, we will point out some major findings to motivate our line of research.

A fundamental assumption underlying today's preference-based information systems posits that a full description of applicable preference statements can be provided together with a query. In contrast, experiments show that the user is not aware of most preferences at query time but rather constructs them dynamically during the choice process. However, these preferences are not arbitrary, but based on individual values, beliefs, and fundamental preferences (Lichtenstein & Slovic, 2006).

But even if we assume all relevant preferences to be known, capturing them by utility functions is still difficult. Prospect theory (Kahneman & Tversky, 1979) goes beyond standard utility theory by accounting for three major psychological effects. Utility values cannot be stated ultimately, but are bound to reference points: no item has an intrinsic utility but only can be assessed relative to some other item. Typical reference points are the user's status quo or some average item. Furthermore, the theory accounts for loss aversion, i.e., losses are valued more heavily than corresponding gains. Finally, utilities are subject to saturation effects. Beyond a certain point gains and losses become less important. In the words of Daniel Bernoulli: 'There is no doubt that a gain of one thousand ducats is more significant to the pauper than to a rich man though both gain the same amount.'

Also similarity between items plays a major role when making choices. Tversky's similarity hypothesis (Tversky, 1972) states that when people are confronted with several items to chose from, the addition of an alternative affects items similar to the added alternative more than those that are dissimilar to it. Current information systems largely ignore this effect favoring ease of computation. In any case, similarity measures resembling human perception are difficult to handle, since they often expose non-metric properties (Tversky, 1977; Tversky & Gati, 1982).

We can thus see that the psychological understanding of human preferences is at odds with the preference models used in today's systems. If so, how to build human-centered preference-based information systems?

A fundamental truth in cognitive psychology recognizes categories and concepts as building blocks of human thought (Murphy, 2002). Basically, a category is a class of entities (e.g., items, actions, states, or properties) grouped with respect to some criterion or rule. All items in a category are considered to be similar. Concepts are mental representations of categories. Due to this dualism both terms are often used interchangeably. In essence, human cognition encodes knowledge in terms of concepts, thus enabling people to describe and understand their environment. Also in information systems concepts provide a convenient means for knowledge representation. But again today's systems sacrifice essential real-world properties of concepts for computational simplicity.

Until the 1970s the classical view of concepts prevailed: categories can be logically defined by a set of necessary and sufficient conditions. But especially the work of psychologist Eleanor Rosch pointed to severe flaws in the classical view (Rosch, 1978). The current state of research can be summarized by four main observations (Hampton, 2006):

- **Vagueness.** There are categories without clear rules defining membership.
- **Opacity.** Even if there is clear a rule for categorization, people often are unable to verbalize it.
- **Typicality.** Within a category, some entities may represent the corresponding concept better than others. If asked, people generally are able to name some typical members.
- **Genericity.** Commonly made statements about a given concept's characteristics are true in a general sense, but do not necessarily extend to all members of the respective category.

This leads to a variety of models largely explaining the above phenomena (Hampton, 1997; Murphy, 2002). Two noteworthy models are the prototype view and the exemplar view. The prototype view posits that concepts are represented by most generic examples, so-called prototypes. Prototypes tend to be abstract and may not be existing entities. In contrast, the exemplar view represents concepts by a set of actually existing entities contained in the category. These entities are either very typical or especially remarkable entities that easily come to mind. In both cases categorization of new entities is performed by comparing to either prototypes or exemplars.

There are only a few studies relating concept models to preference statements in a general way. However, it is known that concepts matter a great deal when people are confronted with new items (Mellers, Schwartz, & Cooke, 1998; Devetag, 1999). Familiar concepts raise expectations about those items, which are often subsequently used as reference points. Furthermore, lacking precise information people tend to favor typical items. This cognitive mechanism also explains phenomena commonly observed in choice situations such as the attraction effect, extremeness aversion, and framing.

For verbalizing statements about conceptual knowledge often so-called generics are used. With generics people express 'most primitive and fundamental generalizations' (Leslie, 2007; Leslie, 2008), e.g. 'Ferraris are fast.' Though these generalizations are not always true in a strict logical sense, they help to quickly and intuitively communicate information essential for grasping the respective concept.

Relating the above insights to models of the mind, Daniel Kahneman's Two Systems view of cognition (Kahneman, 2003) proves helpful. Here cognitive processes are facilitated to some degree by *System 1*, a fast, automatic, effortless lower-level system, and to some degree by *System 2*, a slower, more effortful, rule-based higher-level system. Generally both systems are involved in cognitive processes, but the respective degrees may vary. The psychological phenomena mentioned in the current section can be attributed to System 1, whereas mathematical reasoning as exhibited by today's information systems generally falls into the domain of System 2.

## 4. Integrating System 1 into Information Systems

The above considerations raise the question whether it is justified to base information systems solely on formal decision processes. Many decision theorists claim that a choice can only be good it is guarded by an entirely rational—and therefore completely explainable—decision process. Indeed, internal coherence and logical consistency build a solid foundation. However, psychological research shows that people cannot live up to these standards. Nevertheless, decisions made instinctively with limited cognitive resources often still lead to good results that people have a positive feeling about. In natural problem settings, people exploit environmental properties by applying simple heuristics spanning both cognitive systems. Heuristics are 'rules of thumb,' educated guesses, intuitive judgments, or simply common sense blended with some degree of reasoning; the notion of rationality should account for the boundedness of cognitive resources (Simon, 1956; Gigerenzer, 2006; Todd & Gigerenzer, 2007).

In a nutshell, preference-based information systems can not limit themselves to producing hardly digestible logical explanations, but also have to cater for the positive user experience found in natural decision processes. Still, automated reasoning processes should be used to prevent the user from severe and hard to detect fallacies. But in any case, systems should always aim at providing a cognitive match. This starts at query level by additionally supporting concept-driven preference statements, runs through the entire search process usually involving some user interaction, and ends at the presentation of results in an intuitively understandable way. Throughout the process, the properties of preferences pointed out by cognitive psychology have to be respected.

This idea is not entirely new. Already during the 1980s, when designing today's leading database query language SQL, their creators wanted it to be as close to natural language as possible. Actually, it was designed in a declarative way in a simple syntax following a 'SELECT *items* FROM *source* WHERE *condition applies*' template. Still, all parts of the query have to be provided explicitly. As pointed out above, concepts are fundamental to human cognition. By supporting generics, querying promises to be elevated to the next level. The user does not need to externalize all knowledge, but can formulate preferential queries along the way she/he thinks. Explicit statements are only given in the form of 'landmarks' to guide the system. Based on these landmarks, together with the user's notions, the system can now derive a map of the individual choice context.

Realizing this idea in practical systems requires insights from philosophy. Since algorithms work on the level of System 2, a major task is the simulation of System 1's cognitive processes at System 2 respecting all the processes' properties. Information systems design therefore needs expertise from philosophy, computer science, and psychology, crossing each discipline's borders.

## 5. Outlook and System Design

Currently, we are developing a preference-based information system, which applies the ideas given in this paper. We are planning to use this system as a test bed for future research. Figure 1 depicts the general design of our system, which we will explain in detail using an example from e-commerce: buying a used car.

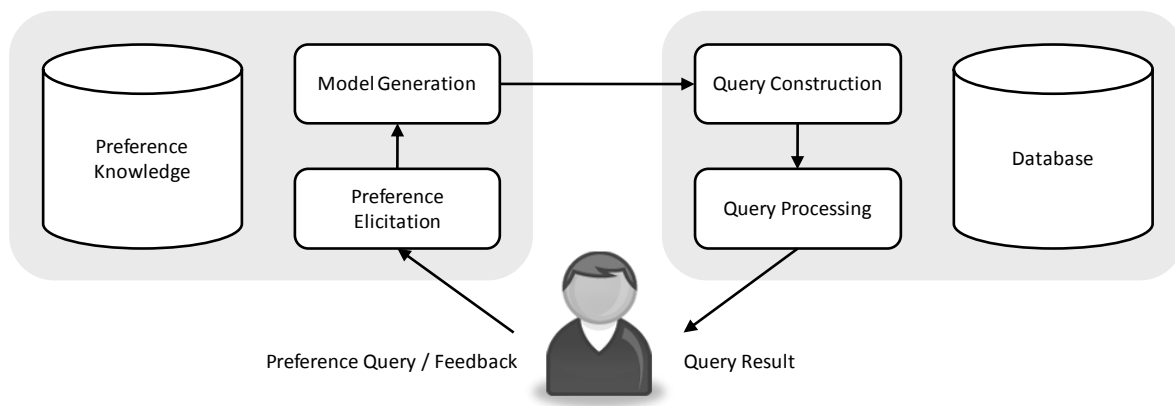

Figure 1: Design of our preference-based information system

Think of a large *database* containing information about used cars offered for sale. Each car is described within a fixed attribute scheme (e.g. make, model, price, age, top speed, and color). In a preprocessing step, conceptual knowledge is automatically extracted from this data, leading to typicality statements like "A typical Ferrari is fast, expensive, and red." Along with many other types of preferential information (e.g. background information about users or sales statistics), these statements are stored in database containing all relevant *preference knowledge* to be used by the system.

A *preference elicitation* module is responsible for the communication with the user and provides an interface to the system as well as a query mechanism. "I want to buy a Ferrari" is an example of a simple *preference query*. Within the system, a *model generation* module translates such statements into a formal representation. This representation is not necessarily based on a single preference model (e.g. some utility-based model) but can also be a blend of several models, which in particular exploit the typicality information generated previously. In our example, the system might derive the following fact: "Since a typical Ferrari has property X and the user did not mention any constraints on X, we can assume that the user likes X." Here, the idea is to fill up missing knowledge about the user's preferences by assuming typical default values, which corresponds to actual human behavior. All facts relevant to the current query then will be integrated by the *query construction* module into a database query, usually expressed in terms of a formal query language (e.g. top-k retrieval functions). Finally, *query processing* takes place, i.e. the previously constructed database query is evaluated over the car database, resulting in a ranked set of *query results,* which are returned to the user. Now, the user is able to inspect this list of recommended cars and give some *feedback* (e.g. by criticizing specific aspects of each recommendation). In an iterative process, this feedback will be analyzed by the system and used to come up with an improved result list. In particular, the user can override the system's preferential assumptions by giving explicit information. In addition to the presentation of results, the system will also provide additional information that might help the user in understanding why she/he got these specific results. In our example, after a few iterations it might turn out that Ferraris are too expensive to be considered by the user. Then the system can recommended a sports car having property X even it is not a Ferrari.

Since we currently are just at the beginning of a large research project, there are more questions than answers. In particular, we are interested in finding answers to the following questions:

1. How do people use conceptual information in actual preferential statements?
2. What conceptual information is contained in the databases to be searched, and how this information can be automatically extracted?
3. How and when can conceptual information integrated into a preference-based data retrieval process?
4. How to build an effective and efficient information system to apply these findings?